\documentclass[10pt]{article}
\usepackage{amssymb}

\usepackage{amsmath}
\usepackage{a4}

\input{tcilatex}

\begin{document}

\title{Scalar and Vector Massive Fields in Lyra's Manifold\thanks{
Presented in the \textit{Fourth International Winter Conference on
Mathematical Methods in Physics}, 9 to 23 August 2004, Rio de Janeiro,
Brazil.}}
\author{R. Casana, C. A. M. de Melo and B. M. Pimentel \\
{\small Instituto de F\'{\i}sica Te\'{o}rica, Universidade Estadual Paulista}%
\\
{\small Rua Pamplona 145, CEP 01405-900, S\~ao Paulo, SP, Brazil}\\
{\small {\ E-mail: \textit{casana@ift.unesp.br, cassius@ift.unesp.br,
pimentel@ift.unesp.br}}}}
\date{}
\maketitle

\begin{abstract}
The problem of coupling between spin and torsion is analysed from a Lyra's
manifold background for scalar and vector massive fields using the
Duffin-Kemmer-Petiau (DKP) theory. We found the propagation of the torsion
is dynamical, and the minimal coupling of DKP field corresponds to a
non-minimal coupling in the standard Klein-Gordon-Fock and Proca approaches.
The origin of this difference in the couplings is discussed in terms of
equivalence by surface terms.
\end{abstract}


\section{Introduction}


After Einstein's approach to gravitation, several others theories have been
developed, as part of efforts to cure problems arising when the
gravitational field is coupled to matter fields. In particular, the problem
of spin coupling to gravitation has a central role in some recent years. The
principal path to incorporate spin in geometrical theories of gravitation is
the use of so called Riemann-Cartan geometry. This geometry has a
nonsymmetric connection, in such a way that a new geometrical concept enters
in scene: the torsion. However, analysing the Cauchy data, one can proof the
torsion is a nonpropagating entity and therefore must be different of zero
only in the interior of matter.

As soon as Einstein presented the General Relativity, Weyl \cite{Weyl}
proposed a new geometry in which a new scalar field accompany the metric
field and change the scale of length measurements. The aim was to unify
gravitation and electromagnetism, but this theory was briefly refuted by
Einstein because the nonmetricity had direct consequences over the spectral
lines of elements which never has been observed.

After some more years, Lyra \cite{Lyra} has proposed a new geometry, with
scalar field for scale changes, that respect the metricity condition. This
theory was developed by Scheibe \cite{Lyra}, Sen \cite{Sen} and several
others as an alternative to describe the gravitational field, and more
recently has been applied to study viscous \cite{ViscLyra} and higher
dimensional \cite{HighdimLyra} cosmological models, domain walls \cite%
{DomWallLyra}, and several others applications. In context of
spin-gravitational coupling, the importance of Lyra's geometry resides in
the fact that the torsion is propagating.

On the other hand, to study the behaviour of scalar and vector massive
fields in non-euclidean manifolds is extremely important in the context of
astroparticle physics and unified theories since a great part of our
knowledge about cosmological data and fundamental interactions is described
by this type of field. A profitable manner of describing these fields is to
use the Duffin-Kemmer-Petiau (DKP) theory. In DKP theory, both particles are
described by only one field with a linear first order differential equation,
very similar to Dirac equation. This similarity can be employed to
facilitate the study of interactions between several fields, just as in
General Relativity \cite{Lunardi 1,ijmpa} and Einstein-Cartan spacetimes %
\cite{spin0,spin1}. However, in the last case is found that DKP theory is
not equivalent to the correspondent Klein-Gordon-Fock (KGF) and Proca
Lagrangians. Notwithstanding, the Harisch-Chandra theory for massless DKP
field \cite{HarishC} was extended to Riemann-Cartan manifold in such a way
that a complete equivalence with KGF and Maxwell theories can be proved \cite%
{CQG}. Therefore, the equivalence between DKP and the more usual theories is
not trivial, and the question of what is the most fundamental theory arises.
Evidently, only a very accurate experiment could decide.

Here, we propose look for coupling of spin 0 and 1 massive fields and
torsion in Lyra manifold via DKP formalism. A good introduction to DKP
theory can be found in \cite{Umezawa,KrajcikNieto}. In section \ref{LyraGeom}
we present the essential elements of Lyra geometry, and in the subsequent
sections the coupling of DKP field with curvature and torsion in this
manifold as well as a comparison between the results of the more usual KGF
and Proca formalisms. Finally, in the last section we make some comments on
the results.


\section{The Lyra Geometry\label{LyraGeom}}


The Lyra manifold \cite{Lyra} is defined given a tensor metric $g_{\mu \nu }$
and a positive definite scalar function $\phi $ which we call the scale
function. In Lyra geometry, one can change scale and coordinate system in an
independent way, to compose what is called a \emph{reference system}
transformation: let $M\subseteq \mathbb{R}^{N}$ and $U$ an open ball in $%
\mathbb{R}^{n}$, ($N\geq n$) and let $\chi :U\curvearrowright M$. The pair $%
\left( \chi ,U\right) $ defines a \emph{coordinate system}. Now, we define a
reference system by $\left( \chi ,U,\phi \right) $ where $\phi $ transforms
like 
\begin{equation*}
\bar{\phi}\left( \bar{x}\right) =\bar{\phi}\left( x\left( \bar{x}\right)
;\phi \left( x\left( \bar{x}\right) \right) \right) \,,\quad \frac{\partial 
\bar{\phi}}{\partial \phi }\not=0
\end{equation*}%
under a reference system transformation.

In the Lyra's manifold, vectors transform as 
\begin{equation*}
\bar{A}^{\nu }=\frac{\bar{\phi}}{\phi }\frac{\partial \bar{x}^{\nu }}{%
\partial x^{\mu }}A^{\mu }
\end{equation*}%
In this geometry, the affine connection is 
\begin{equation}
\tilde{\Gamma}_{\;\,\mu \nu }^{\rho }\equiv \frac{1}{\phi }\mathring{\Gamma}%
^{\rho }{}_{\mu \nu }+\frac{1}{\phi }\left[ \delta _{\,\mu }^{\rho }\partial
_{\nu }\ln \left( \frac{\phi }{\bar{\phi}}\right) -g_{\mu \nu }g^{\rho
\sigma }\partial _{\sigma }\ln \left( \frac{\phi }{\bar{\phi}}\right) \right]
\,,\quad \mathring{\Gamma}_{\,\mu \nu }^{\rho }\equiv \frac{1}{2}g^{\rho
\sigma }\left( \partial _{\mu }g_{\nu \sigma }+\partial _{\nu }g_{\sigma \mu
}-\partial _{\sigma }g_{\mu \nu }\right)  \label{ConexLyra}
\end{equation}%
whose transformation law is given by 
\begin{equation}
\tilde{\Gamma}_{\;\,\mu \nu }^{\rho }=\frac{\bar{\phi}}{\phi }\bar{\Gamma}%
_{\;\,\lambda \varepsilon }^{\sigma }\frac{\partial x^{\rho }}{\partial \bar{%
x}^{\sigma }}\frac{\partial \bar{x}^{\lambda }}{\partial x^{\mu }}\frac{%
\partial \bar{x}^{\varepsilon }}{\partial x^{\nu }}+\frac{1}{\phi }\frac{%
\partial x^{\rho }}{\partial \bar{x}^{\sigma }}\frac{\partial ^{2}\bar{x}%
^{\sigma }}{\partial x^{\mu }\partial x^{\nu }}+\frac{1}{\phi }\delta
_{\,\nu }^{\rho }\frac{\partial }{\partial x^{\mu }}\ln \left( \frac{\bar{%
\phi}}{\phi }\right) \,.  \label{TransConexLyra}
\end{equation}

One can define the covariant derivative for a vector field as 
\begin{equation*}
\nabla _{\mu }A^{\nu }\equiv {\frac{1}{\phi }}\partial _{\mu }A^{\nu }+%
\tilde{\Gamma}_{\;\,\mu \alpha }^{\nu }A^{\alpha }\,,\quad \nabla _{\mu
}A_{\nu }\equiv {\frac{1}{\phi }}\,\partial _{\mu }A_{\nu }-\tilde{\Gamma}%
_{\;\,\mu \nu }^{\alpha }A_{\alpha }\,.
\end{equation*}

The richness of the Lyra's geometry is demonstrated by\ the \emph{curvature} %
\cite{Sen} 
\begin{equation}
\tilde{R}_{\,\beta \alpha \sigma }^{\rho }\equiv \frac{1}{\phi ^{2}}\left( 
\frac{\partial \left( \phi \tilde{\Gamma}_{\,\alpha \sigma }^{\rho }\right) 
}{\partial x^{\beta }}-\frac{\partial \left( \phi \tilde{\Gamma}_{\,\beta
\sigma }^{\rho }\right) }{\partial x^{\alpha }}+\phi \tilde{\Gamma}_{\,\beta
\lambda }^{\rho }\phi \tilde{\Gamma}_{\,\alpha \sigma }^{\lambda }-\phi 
\tilde{\Gamma}_{\,\alpha \lambda }^{\rho }\phi \tilde{\Gamma}_{\,\beta
\sigma }^{\lambda }\right)  \label{CurvLyra}
\end{equation}%
and the \emph{torsion }\cite{DeFelice}

\begin{equation}
\tilde{\tau}_{\mu \nu }^{\,\quad \rho }\equiv \tilde{\Gamma}^{\rho }{}_{\mu
\nu }-\tilde{\Gamma}^{\rho }{}_{\nu \mu }-\frac{1}{\phi }\left( \delta _{\mu
}^{\rho }\partial _{\nu }-\delta _{\nu }^{\rho }\partial _{\mu }\right) \ln
\phi
\end{equation}%
where the second term is the anholonomic contribution, thus, we get 
\begin{equation}
\tilde{\tau}_{\mu \nu }^{\,\quad \rho }=-\frac{1}{\phi }\left( \delta _{\mu
}^{\rho }\partial _{\nu }-\delta _{\nu }^{\rho }\partial _{\mu }\right) \ln 
\bar{\phi}\,,\quad \tilde{\tau}_{\mu }\equiv \tilde{\tau}_{\mu \rho
}^{\,\quad \rho }=\frac{3}{\phi }\partial _{\mu }\ln \bar{\phi}\,.
\end{equation}


\section{The Massive DKP Field in Lyra Manifold}


In Minkowski space--time the massive DKP theory is given by the following
Lagrangian density 
\begin{equation*}
\mathcal{L}\;=\;\frac{i}{2}\overline{\psi }\beta ^{a}\partial _{a}\psi -%
\frac{i}{2}\partial _{a}\overline{\psi }\beta ^{a}\psi -m\overline{\psi }%
\psi ,
\end{equation*}
where $\overline{\psi }=\psi ^{\dagger }\eta ^{0}\,,\;\eta ^{0}=2\left(
\beta ^{0}\right) ^{2}-1\,$, and the $\beta ^{a}$ are matrices satisfying
the massless DKP algebra\footnote{%
We choose a representation in which ${\beta ^{0}}^{\dag }={\beta ^{0}}$, ${%
\beta ^{i}}^{\dag }=-{\beta ^{i}}$\thinspace .} 
\begin{equation*}
\beta ^{a}\beta ^{b}\beta ^{c}+\beta ^{c}\beta ^{b}\beta ^{a}=\beta ^{a}\eta
^{bc}+\beta ^{c}\eta ^{ba}\,.
\end{equation*}

The resulting equation of motion for the DKP field $\psi$ is 
\begin{eqnarray}  \label{eq3}
i\beta^a\partial_a\psi-m\psi=0 \;.
\end{eqnarray}

The above equations can be generalized to Lyra space--time \cite{Lyra} $%
\mathbb{L}$ through the formalism of\emph{\ tetrads} (or \emph{vierbeins})
together the \emph{minimal coupling procedure} \cite{hehl,sabbata}. Here we
shall simply quote the main results we need. For details, in Riemann and
Riemann--Cartan manifolds, we refer respectively to \cite{Lunardi 1,ijmpa}
and \cite{spin0,spin1,CQG} and references therein.

We consider a Lyra space-time $\mathbb{L}$ with metric $g_{\mu \nu }$, whose
point coordinates are labelled $x^{\mu }$. To each point in $\mathbb{L}$ we
attach a Minkowski space-time $\mathbb{M}$ with metric $\eta _{ab}$, whose
point coordinates are labelled $x^{a}$. The DKP fields $\psi $ are \emph{%
Lorentz group} representations in Minkowski space-time. The projections into 
$\mathbb{L}$ of all tensor quantities defined on $\mathbb{M}$ are done \emph{%
via} the tetrad fields $e^{\mu }{}_{a}(x)$ : 
\begin{equation}
g_{\mu \nu }(x)=\eta _{ab}e_{\mu }{}^{a}(x)e_{\nu }{}^{b}(x)\,,\quad e_{\nu
}{}^{a}e^{\nu }{}_{b}=\delta _{\,b}^{a}\,,\quad e=\det \left( e_{\mu
}{}^{a}\right) =\sqrt{-g}\,,  \label{eq4}
\end{equation}%
where $g=\det (g_{\mu \nu })$.

The resulting action for massive DKP fields minimally coupled to Lyra's
manifold is 
\begin{equation}
S_{DKP}=\int d^{4}x\phi ^{4}\,e\left( \frac{i}{2}\,\overline{\psi }\beta
^{a}e^{\mu }{}_{a}\nabla _{\mu }\psi -\frac{i}{2}\,\nabla _{\mu }\overline{%
\psi }e^{\mu }{}_{a}\beta ^{a}\psi -m\overline{\psi }\psi \right) \;,
\label{eq5}
\end{equation}%
where $\nabla _{\mu }$ is the Lyra covariant derivative associated to the
affine connection $\tilde{\Gamma}_{\;\alpha \mu }^{\nu }$.

The covariant derivatives of DKP fields are 
\begin{eqnarray*}
\nabla _{\mu }\psi &=&{\frac{1}{\phi }}\,\partial _{\mu }\psi +\frac{1}{2}%
\omega _{\mu ab}S^{ab}\psi \\
\nabla _{\mu }\overline{\psi } &=&{\frac{1}{\phi }}\,\partial _{\mu }%
\overline{\psi }-\frac{1}{2}\omega _{\mu ab}\overline{\psi }S^{ab}\,,
\end{eqnarray*}
where $S^{ab}=[\beta ^{a},\beta ^{b}]$ and $\omega _{\mu ab}$ is the spin
connection.

The Euler-Lagrange equation for the $\psi $ field is 
\begin{equation}
i\beta ^{\mu }\left( \nabla _{\mu }+\frac{1}{2}\tilde{\tau}_{\mu }\right)
\psi -m\psi =0  \label{eq7linha}
\end{equation}%
where we have used the metricity condition, $\nabla _{\alpha }e_{\mu }^{\,a}=%
\frac{1}{\phi }\partial _{\alpha }e_{\mu }^{\,a}-\tilde{\Gamma}_{\,\alpha
\mu }^{\rho }e_{\rho }^{\,a}+\omega _{\alpha \,b}^{\,a}e_{\mu }^{\,b}\equiv
0 $.


\subsection{The Scalar Sector}


In Minkowski space--time, the ``projectors'' $P$ and $P^{a}$ select the spin
0 sector of the theory (see \cite{Umezawa, Lunardi 1}) such that $P\psi $ is
a scalar and $P^{a}\psi $ is a vector field. Thus, from these projectors
defined in $\mathbb{M}$ we can construct the projectors in Lyra manifold as 
\begin{equation*}
P^{\mu }=e^{\mu }{}_{a}P^{a}=e^{\mu }{}_{a}P\beta ^{a}=P\beta ^{\mu }.
\end{equation*}%
From the definitions above and the properties of $P$ and $P^{a}$ it is easy
to verify that $P^{\mu }\beta ^{\nu }=Pg^{\mu \nu }\,,\ PS^{\mu \nu }=0$,
and it can also be seen that $P\nabla _{\mu }\psi =\nabla _{\mu }\left(
P\psi \right) $ and $P^{\nu }\nabla _{\mu }\psi =\nabla _{\mu }\left( P^{\nu
}\psi \right) $ due to the metricity condition. Therefore, under general
coordinate transformations, $P\psi $ is a scalar and $P^{\nu }\psi $ is a
vector.

By applying the projectors $P$ and $P^{\mu }$ to the equation (\ref{eq7linha}%
), we get respectively,%
\begin{equation}
mP\psi =i\left( \nabla _{\mu }+\frac{1}{2}\tilde{\tau}_{\mu }\right) P^{\mu
}\psi \,,\quad mP^{\mu }\psi =i\left( \nabla ^{\mu }+\frac{1}{2}\tilde{\tau}%
^{\mu }\right) P\psi  \label{eq14}
\end{equation}%
by mixing both equation, we obtain the equation of motion for the scalar $%
P\psi $. We choose a representation where DKP field is a 5-vector column
such as $\psi =(\varphi ,\psi ^{0},\psi ^{1},\psi ^{2},\psi ^{3})^{T}$, $%
P\psi =(\varphi ,0,0,0,0)^{T}$ and $P^{a}\psi =(\psi ^{a},0,0,0,0)^{T}$.
Thus, we have 
\begin{equation}
\left( \nabla _{\mu }+\frac{1}{2}\tilde{\tau}_{\mu }\right) \left( \nabla
^{\mu }+\frac{1}{2}\tilde{\tau}^{\mu }\right) \varphi +m^{2}\varphi =0
\label{eq15}
\end{equation}%
As we can see above, the interaction with torsion does not disappear, even
after we selected the spin 0 sector of the DKP field. This interaction is
present both in the connection $\tilde{\Gamma}_{\;\,\alpha \mu }^{\nu }$
used in the calculation of the covariant derivative $\nabla _{\mu }$ and in
the explicit presence of terms containing the trace torsion $\tilde{\tau}%
_{\mu }$ in the equation above.

On the other hand, when the Lyra geometry is minimally coupled to the
massive Klein-Gordon-Fock field, we get 
\begin{equation}
S_{KG}=\int d^{4}x\;\phi ^{4}\sqrt{-g}\left( \nabla _{\mu }\varphi ^{\ast
}\nabla ^{\mu }\varphi -m^{2}\varphi ^{\ast }\varphi \right) \,,
\label{eq16}
\end{equation}%
where the covariant derivative of the KGF scalar reads $\nabla _{\mu
}\varphi =\frac{1}{\phi }\partial _{\mu }\varphi $.

The KGF action (\ref{eq16}) results in the following equation of motion 
\begin{equation}
\left( \nabla _{\mu }+\tilde{\tau}_{\mu }\right) \nabla ^{\mu }\varphi
+m^{2}\varphi =0\,,  \label{eq18}
\end{equation}%
We can see that there exist interaction with the trace torsion. It is a
different situation to what happened in Riemann-Cartan spacetime where the
scalar field does not couple with torsion \cite{spin0}. However, the spin 0
DKP equation (\ref{eq15}) is different of KGF equation (\ref{eq18}).

The difference will be better understood if we project the DKP action (\ref%
{eq5}) to its spin 0 sector:%
\begin{equation}
\overline{\psi }\beta ^{a}e^{\mu }{}_{a}\nabla _{\mu }\psi =\psi ^{\mu \ast
}\nabla _{\mu }\varphi +\varphi ^{\ast }\nabla _{\mu }\psi ^{\mu }\,,\quad
e^{\mu }{}_{a}\nabla _{\mu }\overline{\psi }\beta ^{a}\psi =\psi ^{\mu
}\nabla _{\mu }\varphi ^{\ast }+\varphi \nabla _{\mu }\psi ^{\mu \ast
}\,,\quad \overline{\psi }\psi =\psi ^{\mu \ast }\psi _{\mu }+\varphi ^{\ast
}\varphi   \label{eq20}
\end{equation}

Thus, the DKP action (\ref{eq5}) for the spin 0 sector, after a rescaling $%
\varphi \rightarrow \sqrt{m}\,\varphi $, reads as 
\begin{equation*}
S_{DKP0}=\int d^{4}x\;\phi ^{4}\sqrt{-g}\left( -\frac{1}{2}\left[ \varphi
^{\ast }\nabla _{\mu }\nabla ^{\mu }\varphi +\varphi \nabla _{\mu }\nabla
^{\mu }\varphi ^{\ast }\right] -m^{2}\varphi ^{\ast }\varphi -\frac{1}{2}%
\nabla _{\mu }\left( \tilde{\tau}^{\mu }\varphi \varphi ^{\ast }\right) -%
\frac{1}{4}\left( \tilde{\tau}^{\mu }\tilde{\tau}_{\mu }\varphi ^{\ast
}\varphi \right) \right) ,
\end{equation*}%
where we have used the equation (\ref{eq14}) to relate the vector $\psi
^{\mu }$ to the scalar $\varphi $.

After some integration by parts, the action (\ref{eq20}) simplifies to read 
\begin{equation*}
S_{DKP0}=\int d^{4}x\;\phi ^{4}\sqrt{-g}\left( \nabla _{\mu }\varphi ^{\ast
}\nabla ^{\mu }\varphi -m^{2}\varphi ^{\ast }\varphi -\frac{1}{2}\nabla
_{\mu }\tilde{\tau}^{\mu }\varphi \varphi ^{\ast }-\frac{1}{4}\tilde{\tau}%
^{\mu }\tilde{\tau}_{\mu }\varphi ^{\ast }\varphi \right) \,,
\end{equation*}
from this action we can obtain the spin 0 DKP equation given in (\ref{eq15}%
). And it has two non minimal coupling which do not appear in the KGF action
(\ref{eq16}).


\subsection{The Vectorial Sector}


Now we use the Umezawa's ``projectors'' $R^{\mu }$ and $R^{\mu \nu }$ in
order to analyze the spin $1$ sector of the theory. We remember that $R^{\mu
}\psi \equiv \psi ^{\mu }$ is a vector and $R^{\mu \nu }\psi \equiv \psi
^{\mu \nu }$ is a second rank antisymmetric tensor in a Lyra sense. Applying
these operators on the equation of motion (\ref{eq3}) we get, respectively,%
\begin{equation}
m\psi ^{\mu }=i\left( \nabla _{\beta }+\frac{1}{2}\tilde{\tau}_{\beta
}\right) \psi ^{\mu \beta }\,,\quad m\psi ^{\mu \beta }=i\left( \nabla
_{\alpha }+\frac{1}{2}\tilde{\tau}_{\alpha }\right) \left( g^{\alpha \beta
}\psi ^{\mu }-g^{\alpha \mu }\psi ^{\beta }\right)  \label{eq24}
\end{equation}%
by mixing both equations, we found the equation of motion for the vector
field $\psi ^{\mu }$ 
\begin{equation*}
\left( \nabla _{\beta }+\frac{1}{2}\tilde{\tau}_{\beta }\right) \left(
\nabla _{\alpha }+\frac{1}{2}\tilde{\tau}_{\alpha }\right) \left( g^{\alpha
\beta }\psi ^{\mu }-g^{\alpha \mu }\psi ^{\beta }\right) +m^{2}\psi ^{\mu }=0
\end{equation*}

We project the massive DKP action (\ref{eq5}) to its spin 1 sector, thus, we
have that each term reads as 
\begin{eqnarray}
\overline{\psi }\beta ^{a}e^{\mu }{}_{a}\nabla _{\mu }\psi &=&\frac{1}{2}%
\psi ^{\ast \mu \nu }\left( \nabla _{\mu }\psi _{\nu }-\nabla _{\nu }\psi
_{\mu }\right) -\psi _{\mu }^{\ast }\nabla _{\nu }\psi ^{\mu \nu }\,,\quad
e^{\mu }{}_{a}\nabla _{\mu }\overline{\psi }\beta ^{a}\psi =\frac{1}{2}\psi
^{\mu \nu }\left( \nabla _{\mu }\psi _{\nu }^{\ast }-\nabla _{\nu }\psi
_{\mu }^{\ast }\right) -\psi _{\mu }\nabla _{\nu }\psi ^{\ast \mu \nu } 
\notag \\
\overline{\psi }\psi &=&-\psi ^{\mu \ast }\psi _{\mu }-\frac{1}{2}\psi
^{\ast \mu \nu }\psi _{\mu \nu }  \label{eq26}
\end{eqnarray}%
and by using the equation (\ref{eq24}) which relates the tensor field $\psi
^{\mu \nu }$ to the vector $\psi ^{\mu }$, $\psi ^{\mu \nu }=-\frac{i}{m}%
\;\left( U^{\mu \nu }+\Sigma ^{\mu \nu }\right) $, where 
\begin{eqnarray}
U_{\mu \nu } &\equiv &\nabla _{\mu }\psi _{\nu }-\nabla _{\nu }\psi _{\mu }=%
\frac{1}{\phi }f_{\mu \nu }+\frac{2}{3}\left( S_{\mu \nu }-\Sigma _{\mu \nu
}\right)  \label{U} \\
f_{\mu \nu } &\equiv &\partial _{\mu }\psi _{\nu }-\partial _{\nu }\psi
_{\mu }\,,\quad \Sigma _{\mu \nu }\equiv \frac{1}{2}\left( \tilde{\tau}_{\mu
}\psi _{\nu }-\tilde{\tau}_{\nu }\psi _{\mu }\right) \,,\quad S_{\mu \nu
}\equiv \frac{3}{2}\frac{1}{\phi }\left( \psi _{\nu }\partial _{\mu }-\psi
_{\mu }\partial _{\nu }\right) \ln \left( \phi \right)  \notag
\end{eqnarray}%
The DKP action (\ref{eq5}) when projected to spin 1 is found to be 
\begin{eqnarray}
S_{DKP1} &=&\int d^{4}x\phi ^{4}e\left( \frac{i}{2}\left[ \frac{1}{2}\psi
^{\ast \mu \nu }\left( \nabla _{\mu }\psi _{\nu }-\nabla _{\nu }\psi _{\mu
}\right) -\psi _{\mu }^{\ast }\nabla _{\nu }\psi ^{\mu \nu }\right] \right) +
\notag \\
&&+\int d^{4}x\phi ^{4}e\left( -\frac{i}{2}\left[ \frac{1}{2}\psi ^{\mu \nu
}\left( \nabla _{\mu }\psi _{\nu }^{\ast }-\nabla _{\nu }\psi _{\mu }^{\ast
}\right) -\psi _{\mu }\nabla _{\nu }\psi ^{\ast \mu \nu }\right] +m\left[
\psi ^{\mu \ast }\psi _{\mu }+\frac{1}{2}\psi ^{\ast \mu \nu }\psi _{\mu \nu
}\right] \right) .  \notag
\end{eqnarray}

We will use the following relation, 
\begin{equation*}
\int d^{4}x\;\phi ^{4}\sqrt{-g}\psi _{\mu }^{\ast }\nabla _{\nu }\psi ^{\mu
\nu }=\int d^{4}x\;\phi ^{4}\sqrt{-g}\left( \frac{1}{2}U_{\mu \nu }^{\ast
}+\Sigma _{\mu \nu }^{\ast }\right) \psi ^{\mu \nu }
\end{equation*}
and after some integration by parts and a rescaling $\psi ^{\mu }\rightarrow 
\sqrt{m}\psi ^{\mu }$, we get the DKP spin 1 action

\begin{eqnarray}
S_{DKP1} &=&\int d^{4}x\phi ^{4}e\left( -\frac{1}{2\phi ^{2}}f^{\mu \nu
}f_{\mu \nu }^{\ast }+m^{2}\psi _{\mu }^{\ast }\psi ^{\mu }-\frac{1}{6\phi }%
\left( f^{\mu \nu }\Sigma _{\mu \nu }^{\ast }+f_{\mu \nu }^{\ast }\Sigma
^{\mu \nu }\right) -\frac{1}{18}\Sigma ^{\mu \nu }\Sigma _{\mu \nu }^{\ast
}+\right.  \label{ActionDKP1} \\
&&\,\ \ \ \ \ \ \ \ \ \ \ \ \ \left. -\frac{1}{2\phi }\frac{2}{3}\left(
f^{\ast \mu \nu }S_{\mu \nu }+S^{\ast \mu \nu }f_{\mu \nu }\right) -\frac{1}{%
6}\frac{2}{3}\left( \Sigma ^{\ast \mu \nu }S_{\mu \nu }+S^{\ast \mu \nu
}\Sigma _{\mu \nu }\right) -\frac{1}{2}\left( \frac{2}{3}\right) ^{2}S^{\ast
\mu \nu }S_{\mu \nu }\right)  \notag
\end{eqnarray}

Otherwise, the Proca's lagrangian in Minkowski space--time is given by 
\begin{equation*}
\mathcal{L}_{PR}=-\frac{1}{2}\left( \partial _{a}A_{b}^{\ast }-\partial
_{b}A_{a}^{\ast }\right) \left( \partial ^{a}A^{b}-\partial ^{b}A^{a}\right)
+m^{2}A_{a}^{\ast }A^{a}
\end{equation*}%
By making the minimal coupling procedure to the Lyra spacetime, we get

\begin{eqnarray}
S_{PR} &=&\int d^{4}x\phi ^{4}e\left( -\frac{1}{2\phi ^{2}}F^{\mu \nu
}F_{\mu \nu }^{\ast }+m^{2}A_{\mu }^{\ast }A^{\mu }+\frac{1}{3\phi }\left(
F^{\mu \nu }\Sigma _{\mu \nu }^{\ast }+F_{\mu \nu }^{\ast }\Sigma ^{\mu \nu
}\right) -\frac{1}{2}\left( \frac{2}{3}\right) ^{2}\Sigma ^{\mu \nu }\Sigma
_{\mu \nu }^{\ast }+\right.  \label{actProc} \\
&&\,\ \ \ \ \ \ \ \ \ \ \ \ \ \ \ \ \ \ \ \ \ \ \left. -\frac{1}{2\phi }%
\frac{2}{3}\left( F^{\ast \mu \nu }S_{\mu \nu }+S^{\ast \mu \nu }F_{\mu \nu
}\right) +\frac{1}{3}\frac{2}{3}\left( \Sigma ^{\ast \mu \nu }S_{\mu \nu
}+S^{\ast \mu \nu }\Sigma _{\mu \nu }\right) -\frac{1}{2}\left( \frac{2}{3}%
\right) ^{2}S^{\ast \mu \nu }S_{\mu \nu }\right)  \notag
\end{eqnarray}%
Observe that making $\Sigma ^{\mu \nu }\rightarrow -\frac{1}{2}\Sigma ^{\mu
\nu }$, $S_{PR}\rightarrow S_{DKP1}$ formally.

\section{Comments}

A simple comparison between the Lagrangians and equations of motion shows us
the unequivalence of DKP theory with KGF and Proca descriptions of scalar
and vector massive particles. However, a more accurated inspection reveals
the spin 0 case as a problem of nonminimal coupling. In the spin 1 case the
situation is more complicated, because all terms in DKP Lagrangian are also
present in the Proca, but with modified coupling constants.

Now, from (\ref{ActionDKP1}) we can see that 
\begin{eqnarray*}
S_{DKP1}^{M_{4}} &=&\int d^{4}x\left( -\frac{1}{2}\left( \partial _{a}\psi
_{b}^{\ast }-\partial _{b}\psi _{a}^{\ast }\right) \left( \partial ^{a}\psi
^{b}-\partial ^{b}\psi ^{a}\right) +m^{2}\psi _{a}^{\ast }\psi ^{a}\right)
\quad \longrightarrow \\
S_{DKP1}^{\mathbb{L}} &=&\int d^{4}x\phi ^{4}e\left( -\frac{1}{2\phi ^{2}}%
f^{\mu \nu }f_{\mu \nu }^{\ast }+m^{2}\psi ^{\mu \ast }\psi _{\mu }-\frac{1}{%
6\phi }\left( f^{\mu \nu }\Sigma _{\mu \nu }^{\ast }+f_{\mu \nu }^{\ast
}\Sigma ^{\mu \nu }\right) -\frac{1}{18}\Sigma ^{\mu \nu }\Sigma _{\mu \nu
}^{\ast }+\right. \\
&&\left. -\frac{1}{3\phi }\left( f^{\ast \mu \nu }S_{\mu \nu }+S^{\ast \mu
\nu }f_{\mu \nu }\right) -\frac{1}{9}\left( \Sigma ^{\ast \mu \nu }S_{\mu
\nu }+S^{\ast \mu \nu }\Sigma _{\mu \nu }\right) -\frac{2}{9}S^{\ast \mu \nu
}S_{\mu \nu }\right)
\end{eqnarray*}%
by the prescription 
\begin{equation}
\partial _{a}\rightarrow D_{\mu }\equiv \nabla _{\mu }+\frac{1}{2}\tau _{\mu
}\,,  \label{NMinPresc}
\end{equation}%
and since the equation (\ref{U}) is valid, one can see that the prescription
(\ref{NMinPresc}) only changes the strength of the coupling. The same
conclusion can be obtained analysing the proportion between the coefficients
of the interactions in equations (\ref{ActionDKP1}) and (\ref{actProc}).

In our future perspectives we will do a study of the relationship between
Lyra geometry and gauge theories, which is now in course using the Utiyama
general theory. At same time, the coupling of Dirac field with this manifold
is in preparation. We hope that these studies can clarify if the
nonequivalence is restricted to manifolds with torsion \emph{and} curvature,
or if it is related to the structure of the field theory used to describe
the particles.

\begin{center}
\textbf{Acknowledgements}
\end{center}

R. C. and C. A. M. M. thank FAPESP (grants 01/12611-7 and 01/12584-0
respectively) for support. B. M. P. thanks CNPq and FAPESP (grant
02/00222-9) for partial support.


\end{document}